\documentclass{article}
\usepackage[accepted]{vietnam} 
\usepackage{natbib}
\usepackage{graphicx}      
\begin{document}
\twocolumn[
\title{HOPS 108: Star-formation triggered by a non-thermal jet?}
\titlerunning{HOPS 108: Star-formation triggered by a non-thermal jet?}
\author{A. K. D\'iaz Rodr\'iguez (; IAA-CSIC), M. Osorio (IAA-CSIC), G. Anglada (IAA-CSIC), S. T. Megeath (U. Toledo)
L. F. Rodr\'iguez (IRyA-UNAM), J. J. Tobin (Leiden), E. Furlan (Caltech), J. F. G\'omez (IAA-CSIC), A. M. Stutz (MPIA),
W. Fischer (NASA/Goddard), P. Manoj (TIFR), B. Gonz\'alez-Garc\'ia (ESAC), T. Stanke (ESO), J. Booker (U. Toledo), B. Ali (SSI)}{akdiaz@iaa.es}

%\address{Instituto de Astrof\'isica de Andaluc\'ia (IAA-CSIC), Glorieta de la Astronom\'ia s/n E-18008 Granada, Spain; email: akdiaz@iaa.es}

\vskip 0.5cm 
]

\begin{abstract}
The nature of the far-IR source HOPS 108 has been a matter of debate in the last years. Previous radio observations detected a 3.6 cm source (VLA 12), coincident with HOPS 108, that was interpreted as a radio jet from this protostar. We present new multi-wavelength (0.7-5 cm), multi-configuration VLA observations as well as archive data (3.6 cm) that reveal VLA 12 as three knots of non-thermal emission, with HOPS 108 close to the central knot. We show that these knots have not been ejected by HOPS 108. We propose that the VLA 12 knots are actually part of a radio jet driven by VLA 11 (HOPS 370), a strong nearby source clearly elongated in the direction of the knots. The position of HOPS 108 in the path of the VLA 11-VLA 12 jet suggests an appealing new scenario: the triggered formation of HOPS 108 by the interaction of the jet with the surrounding medium.
\end{abstract}

\begin{section}{Background}

The brightest far infrared sources in the OMC-2 active star-forming region (420 pc) are HOPS 370 (FIR 3), an intermediate-mass Class I object, and HOPS 108 (FIR 4), a Class 0 protostar \cite{1,2,3}. The two sources have been detected at cm wavelengths \cite{4}: a strong unresolved source
(VLA 11) towards HOPS 370, and an elongated source (VLA 12) towards HOPS 108, which was interpreted as a radio jet from this protostar. The brightest line emission from shock tracers observed in the HOPS survey is found towards HOPS 108 \cite{5}; however, this emission may trace the terminal shock of an outflow from HOPS 370 (see [OI] map \cite{6}).

\end{section}

\begin{figure}
\vskip 0.5cm
\centering
%$\begin{array}{cc}
\includegraphics[angle=0,height=10.cm]{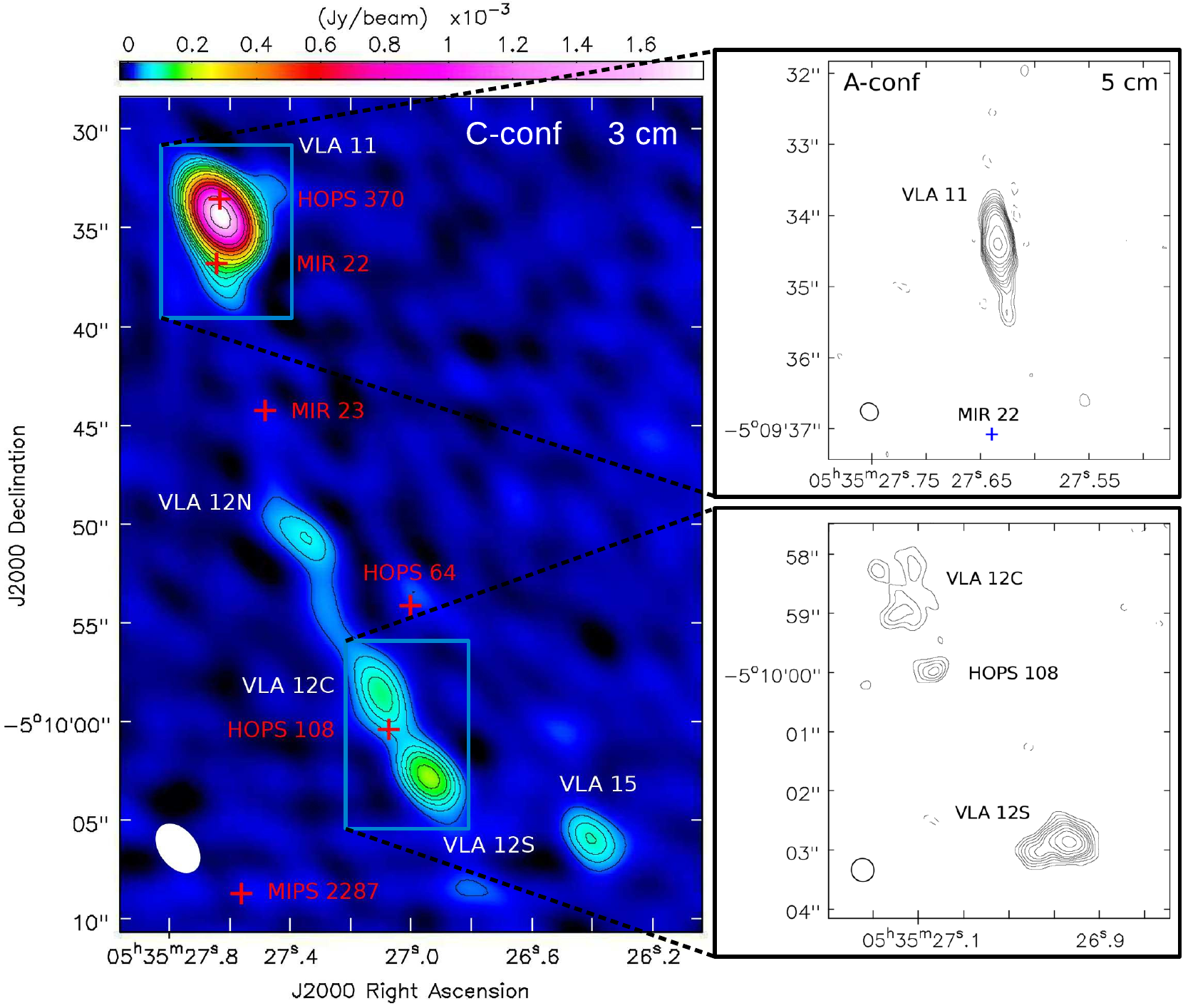} 
%\end{array}$
\caption{{\bf Left}: VLA map at 3 cm of the region around HOPS 108 protostar 
obtained with the C configuration. The 24$\mu$m positions of the MIPS sources\cite{9} are indicated 
by red plus signs. Contour levels are $-$3, 3, 6, 9, 12, 15, 20, 30, 50, 70, 100, 150 and 
200 times the rms noise the map, 9 $\mu$Jy beam$^{-1}$. {\bf Right}: Zoom in at 5 cm obtained with the A configuration. Contour levels are $-$3, 3, 4, 5, 6, 7, 8, 10, 12, 15, 20, 35, 60, 100 and 
150 times the rms noise of each map, 7.5 $\mu$Jy beam$^{-1}$ (top) and 3.2 $\mu$Jy beam$^{-1}$ (bottom). 
The synthesized beam is shown at the bottom left corner of each map. All maps in the figure are adapted\cite{7}.}
\label{fig:X-band}
\end{figure}

\begin{section}{Observations and Results}

We analyze new (2014-2015) multi-wavelength (0.7-5 cm), multi-configuration VLA observations as well as archive (1990-2000) data at 3.6 cm. In figure \ref{fig:X-band} (adapted from \cite{7}) we present the 3 cm map and the high resolution 5 cm map. VLA 11 (HOPS 370) appears clearly elongated (Fig. \ref{fig:X-band}, top right panel), tracing the base of a collimated radio jet; VLA 12 is resolved into three knots (VLA 12N, 12C, 12S, see Fig. \ref{fig:X-band}, left panel), with HOPS 108 close to the central knot (Fig. \ref{fig:X-band}, bottom right panel). At higher frequencies (1.3-0.7 cm) we detect an unresolved source coincident in position with HOPS 108. We detect at all bands a new source that we named VLA 15. The sources VLA 11 (HOPS 370), HOPS 108 and VLA 15 have positive spectral indices, indicating thermal free-free emission from protostars. The VLA 12 knots have negative spectral indices suggesting synchrotron emission due to relativistic electrons accelerated in jet shocks. In a range of $\sim$ 20 years, VLA 11 (HOPS 370), HOPS 108, and VLA 15 remain stationary. We interpret these sources as free-free radio emission originated at the position of embedded protostars. On the other hand, the VLA 12 knots move away from VLA 11 (HOPS 370), in a direction close to the elongation of this source. VLA 12N and 12C move towards HOPS 108, discarding this object as a possible origin of the VLA 12 knots. Since HOPS 108 falls in the path of the VLA 11-VLA 12 jet, and it is associated with signatures of strong shocks, we propose that this protostar could be the result of triggered star formation.

\end{section}

\begin{section}{Conclusions}

Summarizing, VLA 11 traces the thermal region closest to the origin of a collimated radio jet driven by the HOPS 370 intermediate-mass protostar. The knots of the VLA 12 source move away from HOPS 370, present non-thermal emission and are associated with shock tracers, suggesting that they are a non-thermal (synchrotron) lobe of the HOPS 370 jet. The HOPS 108 class 0 protostar is discarded as the origin of the VLA 12 knots. Instead, it’s formation could be triggered \cite{8} by the interaction of the jet with the surrounding medium.

\end{section}

\begin{thebibliography}

\bibitem[Adams(2012)]{1} Adams, J.~D., Herter, T.~L., Osorio, M., et al., {2012}, {Apjl},  {749}, {L24}.

\bibitem[Furlan(2014)]{2} Furlan, E., Megeath, S.~T., Osorio, M., et al.,  {2014}, {ApJ}, {786}, {26}.
%
\bibitem[Furlan(2016)]{3} Furlan, E., Fischer, W.~J., Ali, B., et al., {2016}, {ApJS}, {224}, {5}.
%%
\bibitem[Reipurth(1999)]{4} Reipurth, B., Rodr{\'i}guez, L.~F., \& Chini, R., {1999}, {AJ}, {118}, {983}.
%%
\bibitem[Manoj(2013)]{5} Manoj, P., Watson, D.~M., Neufeld, D.~A., et al., {2013}, {ApJ}, {763}, {83}.
%%
\bibitem[Gonzalez-Garcia(2016)]{6} Gonz{\'a}lez-Garc{\'i}a, B., et al., {2016}, {A\&A}, {596}, {A26}.
%%
\bibitem[Osorio(2017)]{7} Osorio, M., D\'iaz Rodr\'iguez, A. K., Anglada, G., et al., 2017, ApJ, submitted.
%%
\bibitem[Shimajiri(2008)]{8} Shimajiri, Y., Takahashi, S., Takakuwa, S., Saito, M., \& Kawabe, R., {2008}, {ApJ}, {683}, {255}.
%%
\bibitem[Megeath(2012)]{9} Megeath, S.~T., Gutermuth, R., Muzerolle, J., et al., {2012}, {AJ}{144}, {192}.

\end{thebibliography}
\end{document}